# An Integration of General Relativity with Quantum Theory and the Standard Model I

**Joseph E. Johnson, PhD**
**Department of Physics and Astronomy**
**University of South Carolina**
**Columbia, South Carolina, 29208, USA**
**March 7, 2017**

**Abstract**

We propose (1) that the flat space-time metric that defines the traditional covariant Heisenberg algebra commutation rules of quantum theory between the four-vector position and momentum, be generalized to be the space-time dependent Riemann metric following Einstein's equations for general relativity, which determine the metric from the energy-momentum tensor. The metric is then a function of the four-vector position operators which are to be expressed in the position representation. This then allows one (2) to recast the Christoffel, Riemann, and Ricci tensors and Einstein's GR differential equations for the metric as an algebra of commutation relations among the four-vector position and momentum operators (a generalized Lie algebra). This then allows one (3) to generalize the structure constants of the rest of the Poincare algebra with the space-time dependent metric of general relativity tightly integrating it with quantum theory. Finally, (4) we propose that the four-mometum operator be generalized (to be gauge covariant) to include the intermediate vector bosons of the standard model further generalizing this algebra of observables to include gauge observables. Then the generalized Poincare algebra, extended with a four-vector position operator, and the phenomenological operators of the non-Abelian gauge transformations of the standard model form a larger algebra of observables thus tightly integrating all three domains. Ways in which this may lead to observable effects are discussed.

1. **Introduction**

The integration of Einstein's General Relativity (GR) with Quantum Theory (QT) and the (evolving) Standard Model (SM) has frustrated diverse attempts over the last hundred years although each of these three theoretical structures has proved their separate validity beyond question in their respective domains of applicability. Generally speaking, QT[1,2] and the SM[3,4] are founded upon the theory of representations of an algebra of observables. More precisely, elementary particles are each described by distinct representations of the ten parameter Poincare Lie Algebra (PA) of observables ($M^{\mu\nu}$, the Lorentz algebra of three rotations and three Lorentz transformations, along with the four-momentum operators $P^\mu$, that generate translations in space-time) and along with the Heisenberg Lie Algebra (HA) of $P^\mu$ (E/c, $P^i$) and $X^\mu$ (ct, $X^i$) forming the foundation of QT. The SM of non-Abelian Yang-Mills gauge transformations presents another Lie algebra but of non-space-time observables, the SU(3) x SU(2) x U(1) gauge group, that restrict the admissible PA representations, define the allowable interactions among the particles, and provide a theory of dynamical evolution. The representations of a maximum set of commuting elements of these algebras are used to index the representation space, $|\alpha_1, \alpha_2, ...\rangle$ of particle states and to index operators a and $a^+$ that annihilate and create these particles (representations) when acting on the vacuum so that $|\alpha_1, \alpha_2, ...; \beta_1, \beta_2 ...\rangle$ = $\mathbf{a^+}_{\alpha_1, \alpha_2, ...} \mathbf{a^+}_{\beta_1, \beta_2, ...} |0\rangle$. The additional discrete group of transformations of space inversion $I_S$, time inversion, $I_T$, and particle conjugation, $I_C$, are also well defined observables in term of their commutation rules with $M^{\mu\nu}$, $P^\mu$, and $X^\mu$. The transformations that interchange identical particles, **a** and **a⁺**, in symmetric (for integer spin) and antisymmetric (for half integer spin) states are incorporated in the commutation or anticommutation relations that follow from the connection between spin and statistics thus forming an algebraic structure under the permutation group.

But GR[5,6,7] lives in a totally different mathematical environment where there are no "operators" representing observables and thus no commutation rules to define an associated operator algebra. Thus there are no algebraic representations to define the theory as there are in QT. GR is instead founded on non-linear differential equations for the Riemannian curved metric of space-time as determined by the energy-momentum tensor. The "observables" in GR are the four vector positions of events and the metric which defines the distance between two events. The Christoffel symbols, with the Riemannian and Ricci tensors are derivatives of that metric and describe the force of gravity as geodesic paths in this four dimensional space as formed from derivatives of that metric. Einstein's theory of general relativity (GR) can normally be ignored in the quantum physics at small scales as described by the SM but it dominates the large scale structure of the universe. In GR the metric of space-time is curved by the presence of matter and energy as given by $R_{\alpha\beta}$ - ½ $g_{\alpha\beta}$ R + $g_{\alpha\beta}\Lambda$ = (8 π G/c⁴) $T_{\alpha\beta}$ where $R_{\alpha\beta}$ is the Ricci tensor and $T_{\alpha\beta}$ is the energy-momentum tensor expressed in terms of the fundamental particles in the SM theory (along with Dark Matter (DM)). $\Lambda$ represents a constant value that could possibly represent the expansion of the universe associated with Dark Energy (DE). The Ricci and Riemann tensors are defined in terms of the derivatives of the Christoffel symbols which are in turn defined in terms of derivatives of the metric tensor. Attempts to integrate GR with the SM using a gauge group with a massless spin two particle (graviton) have not been successful. But recalling Einstein's basic concepts, gravitation was not to be thought of as a force but rather the consequence of a curvature of space-time induced by the energy-momentum tensor.

In order integrate GR with QT as an algebra of observables, we must identify the primary observables in GR which are the four position vector of events and the metric which is a function of this four dimensional space time. The foundation of QT is the HA which already represents the four vector of position as an operator and also has a Euclidian metric as the structure constants which could be generalized to be functions of the position operators as is the metric in GR. Also, the representation of the HA four momentum operator in the position representation is the derivative which leads to the differential equations in QT. Thus the differential equations of GR could be recast as commutators with the four momentum operators to lead to the algebraic form of Einstein's equations that we seek. Thus we will generalize the flat space-time metric in the HA to be a function of the four position operators in the position representation, where the HA metric "structure constants" are to be the solutions to Einstein's equations derived from the generalized algebraic form of the HA. This generalization can then be extended to the structure constants of the Poincare algebra via the definition of the angular momentum four tensor defined from P and X. We will first extend the PA of the $M^{\mu\nu}$ and $P^{\mu}$ observables to include the four-position operators, $X^{\mu}$ from the HA. We call this combined M, P, X algebra the Extended Poincare (EP) algebra as described in previous work by the author [8].

2. **The Extended Poincare (EP) Algebra**

As quantum theory is founded upon the relationship between momentum and position operations as defined in the HA, with [X, P] = iħ, and [E, t] = iħ, then a full Lie algebra of space-time observables must also include a four-position operator $X^{\mu}$ in order to formally include the foundations of quantum theory. This led us previously to extend the PA by adjoining a four-vector position operator, $X^{\mu}$ ($\mu$, $\nu$, ... = 0, 1, 2, 3) whose components are to be considered as fundamental observables using a manifestly covariant form of the HA. As the $X^{\mu}$ generate translations in momentum, they do not generate symmetry transformations or represent conserved quantities but do provide the critical observables of space-time. We choose the Minkowski metric $g^{\mu\mu}$ = (+1, -1, -1, -1) and write the HA in the covariant form as $[P^{\mu}, X^{\nu}]$ = iħ $g^{\mu\nu}$ I where I is an operator that commutes with all elements and has the unique eigenvalue "1" with $P^0$ =E/c, $P^1$= $P_x$ ..., $X^0$ = ct, $X^1$.= x ... . Here "I" is needed to make the fifteen (15) fundamental observables in this EP algebra close into a Lie algebra with the structure constants as follows:

[I, $P^{\mu}$] = [I, $X^{\nu}$] = [I, $M^{\mu\nu}$] = 0 thus I commutes with all operators and has "1" as the only eigenvalue. (1a)
$[P^{\mu}, X^{\nu}]$ = iħ $g^{\mu\nu}$ I which is the covariant Heisenberg Lie algebra – the foundation of quantum theory. (1b)
$[P^{\mu}, P^{\nu}]$ = 0 which insures noninterference of energy momentum measurements in all four dimensions. (1c)
$[X^{\mu}, X^{\nu}]$ = 0 which insures noninterference of time and position measurements in all four dimensions. (1d)
$[M^{\mu\nu}, P^{\lambda}]$ = iħ ($g^{\lambda\nu} P^{\mu} - g^{\lambda\mu} P^{\nu}$) which guarantees that $P^{\lambda}$ transforms as a vector under $M^{\mu\nu}$. (1e)
$[M^{\mu\nu}, X^{\lambda}]$ = iħ ($g^{\lambda\nu} X^{\mu} - g^{\lambda\mu} X^{\nu}$) which guarantees that $X^{\lambda}$ transforms as a vector under $M^{\mu\nu}$. (1f)
$[M^{\mu\nu}, M^{\rho\sigma}]$ = iħ ($g^{\mu\sigma}M^{\nu\rho} + g^{\nu\rho}M^{\mu\sigma} - g^{\mu\rho}M^{\nu\sigma} - g^{\nu\sigma}M^{\mu\rho}$). (1g)
which guarantees that $M^{\rho\sigma}$ transforms as a tensor under the Lorentz group generated by $M^{\mu\nu}$.

The representations of the Lorentz algebra are well-known [8, 9] and are straight forward but the extension to include the four-momentum with the Poincare algebra representations are rather messy. But with our extension of the Poincare algebra to include a four-position operator, the representations are clearer. Because $X^{\mu}$ is now in the algebra, one can now define the orbital angular momentum four-tensor, operator $L^{\mu\nu}$ as:

$L^{\mu\nu} = X^{\mu} P^{\nu} – X^{\nu} P^{\mu}$ (2a)

From this it follows that

$[L^{\mu\nu}, P^{\lambda}]$ = iħ ($g^{\lambda\nu} P^{\mu} - g^{\lambda\mu} P^{\nu}$) (2b)
$[L^{\mu\nu}, X^{\lambda}]$ = iħ ($g^{\lambda\nu} X^{\mu} - g^{\lambda\mu} X^{\nu}$) (2c)
$[L^{\mu\nu}, L^{\rho\sigma}]$ = iħ ($g^{\mu\sigma} L^{\nu\rho} + g^{\nu\rho} L^{\mu\sigma} - g^{\mu\rho} L^{\nu\sigma} - g^{\nu\sigma} L^{\mu\rho}$). (2d)

One can then define an intrinsic spin four-tensor as:

$S^{\mu\nu} = M^{\mu\nu} - L^{\mu\nu}$ (3a)

with the result that

$[S^{\mu\nu}, P^{\lambda}]$ = 0 (3b)
$[S^{\mu\nu}, X^{\lambda}]$ = 0 (3c)
$[S^{\mu\nu}, L^{\rho\sigma}]$ = 0 (3d)
$[S^{\mu\nu}, S^{\rho\sigma}]$ = iħ ($g^{\mu\sigma} S^{\nu\rho} + g^{\nu\rho} S^{\mu\sigma} - g^{\mu\rho} S^{\nu\sigma} - g^{\nu\sigma} S^{\mu\rho}$) (3e)

Now one can separate this EP algebra into the product of two Lie algebras, the nine parameter HA (consisting of X,P,I) and the six parameter homogeneous Lorentz algebra (consisting of the $S^{\mu\nu}$). Thus one can write all EP representations as products of the representations of the two algebras. For the HA one can choose the position representation:

$X^{\mu}|y> = y^{\mu}|y>$  or the momentum representation (4a)

$P^{\mu}|k> = k^{\mu}|k>$ (4b)

or equivalently diagonalize the mass and the sign of the energy and three momenta as

$P^{\mu}P_{\mu} = m^2$, $\varepsilon(P^0)$, with eigenstates written as $|m, \varepsilon(P^0), k>$ (4c)

All representations of the homogeneous Lorentz group have been found by Bergmann and by Gelfand, Neimark, and Shapiro [2,9] to be given by the two Casimir operators $b_0$ and $b_1$ defined as:

$b_0^2 + b_1^2 - 1 = \frac{1}{2} g_{\mu\rho} g_{\nu\sigma} S^{\mu\nu} S^{\rho\sigma}$ (5a)

where $b_0 = 0, \frac{1}{2}, 1, 3/2, ...(|b_1|-1)$ and where $b_1$ is a complex number

$b_0 b_1 = -\frac{1}{4} \varepsilon_{\mu\nu\rho\sigma} S^{\mu\nu} S^{\rho\sigma}$ (5b)

with the rotation Casimir operator as $S^2$ which has the spectrum $s(s+1)$ with the total spin

$s = b_0, b_0+1, ..., (|b_1| - 1)$ (5c)

and the z component of spin:

$\sigma = -s, -s+1, ....s-1, s$ (5d)

Thus the homogeneous Lorentz algebra representation can be written as $|b_0, b_1, s, \sigma>$ which joined with the Heisenberg algebra gives the full representation space as either

$|k^{\mu}, b_0, b_1, s, \sigma> = a^+_{k, b0, b1, s, \sigma}|0>$ (6)

for the momentum representation or

$|y^{\mu}, b_0, b_1, s, \sigma> = a^+_{y, b0, b1, s, \sigma}|0>$ (7)

for the position representation.

These simultaneous eigenvalues represent a maximal set of commuting observables for the EP Lie algebra that can be used to index creation and annihilation operators representing particles (fields) with the quantum numbers shown. The fundamental particles must be in the representation space of these operators and their Lie algebra. Stated another way, the algebra of observable operators must have the particles that exist in nature as the representation space of that algebra.

### 3. The Standard Model (SM)

But while the known elementary particle states can easily be fit into this infinite array of spins and continuous masses, one has a vast overabundance of states as well as a lack of a dynamical theory of their interactions. One would like to have an algebraic structure that gives all possible particles and only those particles as representations. It is here that one imposes the additional requirements of the phenomenological Standard Model (SM) which only allows three Lorentz representations (specified by $b_0$ and $b_1$) which are the pairs of values (½, ±3/2), (0,1) and (0,2) along with a certain spectrum of masses. Specifically these representations of the Lorentz algebra for particles in the SM are as follows: (a) a unique Dirac spin s = ½ for quarks and leptons which is given by $b_0 = ½$ and, $b_1 = ±3/2$ where the sign of $b_1$, $\varepsilon(b_1)$, distinguishes the representation from the conjugate representation and thus where the four states of $|b_0, b_1, s, \sigma>$ can be abbreviated as $|\varepsilon(b_1), \sigma =± ½>$. These four (spinor) states support the definition of the $\gamma^{\mu}$ matrices which result from the requirement that both representation and conjugate representation be used in order for the state to be invariant under a spatial reflection which takes one from the representation to the conjugate representation having the opposite sign of $b_1$ thus giving the standard Dirac theory. When $b_0 = 0$ then the representation is equivalent to its own conjugate and thus one does not have $\varepsilon(b_1)$. There are two such pertinent cases for bosons in the SM: (b) a unique spin

s = 0 (e.g. the Higgs) which is given by $b_0 = 0$ and, $b_1 = 1$, and (c) the four-vector representation given by $b_0 = 0$ and, $b_1 = 2$, which gives both s = 0 and s = 1. Linear combinations of these four spin states ( s=0, σ=0 and s = 1, σ= +1, 0 , −1) can be used to form a four-vector representation which is needed for the photon (electromagnetic potential $A^\mu$ ) and the W and Z vector fields as well as the gluons. The dynamical theory is introduced via the phenomenological standard model (SM) which imposes the requirement that these representations also support the SU(3) x SU(2) x U(1) gauge group which mixes the observables contained in the EP algebra ($X^\mu$, $P^\nu$, $\gamma^\mu$, s, σ) with new gauge observables (electric charge, hypercharge, isospin, color, and flavor which currently lack a space-time origin other than position dependent phase) to account for the masses and spins of physical particles along with their strong and electroweak interactions. This is similar to classical mechanics where the descriptive kinematical infrastructure supports the separate dynamical theory which contains the interactions and a specification of which physical states (masses, charges, magnetic moments…) occur in nature.

### 4. Proposed Method of Integrating GR with QT:

We seek a formulation of GR as an algebra of observables in keeping with the framework of QT. We know that the "observables" in GR are invariant lengths determined by the Riemann metric $g^{\mu\upsilon}$ ($X^\lambda$) and its derivatives as functions of four-position. So the critical observable operators must be (analytic functions of) the position operators $X^\nu$ from the HA, where $[P^\mu, X^\nu] = i\hbar g^{\mu\upsilon}$. Then the Euclidian metric that gives the HA structure constants would be generalized to be $g^{\mu\upsilon}(X^\lambda)$ written as a function of these operators. The representations of the four momentum $P^\mu$ in the Heisenberg Lie algebra in the position representation are the partial derivatives with respect to $X^\lambda$. So if we generalize the metric in the Heisenberg algebra to be the Riemann metric in Einstein's equations and express the algebra in the position representation, then the Christoffel symbols, Ricci, and Riemann tensors which are all derivatives of the metric can be rewritten as commutation relations satisfying our objective as shown below. Since the Heisenberg algebra is the foundation of QT, this would tightly integrate GR and QT into a new algebra in which the structure constants are now the Riemann metric. For this to work, we must also require that $[X^\mu, X^\nu] = 0$ to allow simultaneous measurement of the components of space-time.

Our first motivation was that if one is to integrate general relativity with quantum theory then it must be integrated with the fundamental HA whose structure constants were explicitly given in terms of the metric of flat space-time as well as with the PA. Our second motivation is that if one thinks of these commutation relations as representing the observational interference of measurements of position and momentum then it seems natural that such interference would be altered in a curved space-time and dependent upon the metric that gives the curvature. Our third motivation is to follow the original design of Sophius Lie in the invention of Lie algebras to create a more powerful framework for differential equations by recasting Einstein's GR equations in the form of operator commutation relations in line with QT and the SM. Here this will be possible since in the position representation, the four-momentum operator commutators have representations in a Hilbert space as derivatives thus allowing us to convert all derivatives of the metric to algebraic commutation relations. Our fourth motivation is to treat gravitation more as Einstein envisioned as a modification of our space-time structure as opposed to a force as is normally done by integrating it with the Standard Model (SM) with a graviton. Specifically, we propose a generalization of the flat space time metric in the EP algebraic structure constants to be the Einstein metric for curved space-time (in a consistent manner) thereby integrating the Riemann space time structure with the fundamental position and momentum operators and the EP algebra. We call this generalized EP the "Extended Poincare Einstein" (EPE) algebra where the Riemann metric becomes the structure constants..

However, the EPE is not formally a Lie algebra since (1) the structure constants now vary in space-time and are operators. (2) Secondly because $g^{\mu\upsilon}$ (X) is not a member of the algebra but a function of the X operators, it has to be thought of as an analytic function of the representations of $X^\nu$ in the enveloping algebra. (3) The resulting commutators give expressions involving both the space-time dependent metric and its derivatives. The energy-momentum tensor $T_{\alpha\beta}$ is to be determined from the SM and QT. But due to the relative weakness of the gravitational interactions, $T_{\alpha\beta}$ could be approximated by a classical solution to Einstein's equations such as the Schwarzschild solution which could determine the metric in a small local domain near a massive spherical body. That Schwarzschild metric would then determine the local structure constants and allow one to seek the EPE Lie algebra representations in that local domain using that (locally constant) metric for the structure constants. These representations would then determine the allowable fields and the associated equations such as the modified Dirac equation for the electron in a hydrogen atom, a harmonic oscillator, and a proposed generalized uncertainty principle as described below. One of these could lead to observational tests. In essence we are seeking an expression of the fundamental equations of physics in terms of observables as operators which are defined by commutation rules and whose representations provide the observable particle fields and associated mathematical defining structure and dynamics.

We postulate (1) that the position operators commute with noninterfering simultaneous measurement:

$[X^\mu, X^\nu] = 0$ with real eigenvalues $y^\mu$ on the eigenvectors $|y>$ with notation $X^\mu | y > = y^\mu | y >$ (8a)

and (2) that the Heisenberg Lie algebra structure constants are now given by the Riemann metric

$[P^\mu, X^\nu] = i\hbar \, I \, g^{\mu\nu}(X)$ so we can now also write (8b)

$g^{\mu\nu}(X) = (-i/\hbar) [P^\mu, X^\nu]$ (8c)

implying that the metric tensor $g^{\mu\nu}(X)$ is also determined by the positon-momentum commutator as well as from Einstein's equations from the metric tensor. In the position representation one now has

$<y|P^\mu|\Psi> = (i\hbar \, g^{\mu\nu}(y) \, (\partial/\partial y^\nu) + f^\mu(y)) \, \Psi(y) = (i\hbar \, \partial^\mu + f^\mu(y)) \, \Psi(y)$ (8d)

where $\Psi(y) = <y|\Psi>$ and $\partial^\mu = g^{\mu\nu}(y) \, (\partial/\partial y^\nu)$ and $f^\mu(y)$ is an yet undetermined vector function of $X^\nu$ (8e)

It follows that $[P^\mu, [P^\nu, X^\rho]] \neq 0$ so that the Heisenberg algebra is no longer nilpotent. Instead one gets

$<y| [P^\mu, g^{\alpha\beta}] = i\hbar \, g^{\mu\nu} \, (\partial g^{\alpha\beta}/\partial y_\nu) <y| = i\hbar \, \partial^\mu g^{\alpha\beta}(y) <y|$ since $[f^\mu, g^{\alpha\beta}] = 0$ as they both are functions of X (8f)

From now on $g^{\alpha\beta} = g^{\alpha\beta}(y)$ is to be understood. Thus in the position representation one can write

$g^{\mu\nu} (\partial/\partial y^\nu) f(y) = \partial^\mu f(y) = -(i/\hbar) [P^\mu, f(y)]$ (8g)

for any function $f(y)$ thus allowing one to convert differential operations into commutators with $P^\mu$.

If we had chosen to use the Riemann covariant derivative, then since the covariant derivative of the metric tensor is zero, one could not express the Christoffel symbols or the Riemann and Ricci tensors in terms of that representation of the momentum but rather it must be expressed as the ordinary derivative in order to convert the sequence of derivatives to commutator form.

But the momentum commutators are now no longer zero but in the position representation instead give

$<y| [P^\mu, P^\nu] = [(i\hbar \, g^{\mu\alpha}(y) \, (\partial/\partial y^\alpha) + f^\mu(y)), (i\hbar \, g^{\nu\beta}(y) \, (\partial/\partial y^\beta) + f^\nu(y))] <y|$ or (8h)

$<y| [P^\mu, P^\nu] = (-\hbar^2 (g^{\mu\alpha}(y) \, (\partial g^{\nu\beta}(y)/\partial y^\alpha) \, (\partial/\partial y^\beta) - g^{\nu\alpha}(y) \, (\partial g^{\mu\beta}(y)/\partial y^\alpha) \, (\partial/\partial y^\beta)$
$+ g^{\mu\alpha}(y) \, g^{\nu\beta}(y) \, (\partial/\partial y^\alpha) \, (\partial/\partial y^\beta) - g^{\nu\alpha}(y) \, g^{\mu\beta}(y) \, (\partial/\partial y^\alpha) \, (\partial/\partial y^\beta)) + [P^\mu, f^\nu]) <y|$ (8i)

The third and fourth terms cancel allowing one to re-express the momentum commutator as

$<y| [P^\mu, P^\nu] = (-\hbar^2 (g^{\mu\alpha}(y) \, (\partial g^{\nu\beta}(y)/\partial y^\alpha) - g^{\nu\alpha}(y) \, (\partial g^{\mu\beta}(y)/\partial y^\alpha)) \, (\partial/\partial y^\beta) + [P^\mu, f^\nu]) <y|$ (8j)

One can express $(\partial/\partial y^\beta) = -(i/\hbar) P_\beta$ to get (8k)

$[P^\mu, P^\nu] <y| = (i\hbar \, B^{\mu\nu\beta} \, P_\beta + [P^\mu, f^\nu]) <y| = (i\hbar \, B^{\mu\nu}{}_\beta \, P^\beta + [P^\mu, f^\nu]) <y|$ (8l)

But since this is true on all states $<y|$, It follows that

$[P^\mu, P^\nu] = i\hbar \, B^{\mu\nu}{}_\gamma \, P^\gamma + [P^\mu, f^\nu]$ where we define (8m)

$B^{\mu\nu}{}_\gamma <y| = (g^{\mu\alpha}(y) \, (\partial g^{\nu\beta}(y)/\partial y^\alpha) - g^{\nu\alpha}(y) \, (\partial g^{\mu\beta}(y)/\partial y^\alpha)) \, g_{\beta\gamma}(y) <y|$ (8n)

where the "structure constants" depend upon the both the metric and its derivatives.

This noncommutative four-momentum does not affect the re-expression of Einstein's equations for the metric as commutators, but does affect most of the other EPE commutators such as for angular momentum and spin. Non-commutativity of $[P^\mu, P^\nu]$ is already familiar in the SM for the effective momentum.

The term $[P^\mu, f^\nu]$, where $f^\nu$ is at this point an arbitrary function of the position operators, is a derivative of a vector function of position and thus is a function of position which commutes with the metric and does not alter the commutators of the Christoffel, Riemann, and Ricci tensors. The remaining EP commutators involving angular momentum and spin need to now be re-computed using these results.

We now seek to cast the LHS of the Einstein equations into the form of commutators of algebraic observables. Although the results look complex, they are no more so than the differential equations for the Christoffel, Riemann, and Ricci tensors. The Christoffel symbols

$$\Gamma_{\gamma\alpha\beta} = (½)(\partial_\beta, g_{\gamma\alpha} + \partial_\alpha, g_{\gamma\beta} - \partial_\gamma, g_{\alpha\beta}) \tag{9a}$$

can be written in terms of the commutators of the four-momentum with the metric as

$$\Gamma_{\gamma\alpha\beta} = (½)(-i/\hbar)([P_\beta, g_{\gamma\alpha}] + [P_\alpha, g_{\gamma\beta}] - [P_\gamma, g_{\alpha\beta}]) \tag{9b}$$

Then using $g_{\alpha\beta}(X) = (-i/\hbar)[P_\alpha, X_\beta]$ one obtains

$$\Gamma_{\gamma\alpha\beta} = (-½)(1/\hbar^2)([P_\beta, [P_\gamma, X_\alpha]] + [P_\alpha, [P_\gamma, X_\beta]] - [P_\gamma, [P_\alpha, X_\beta]]) \tag{9c}$$

The Riemann tensor becomes:

$$R_{\lambda\alpha\beta\gamma} = (-i/\hbar)([P_\beta, \Gamma_{\lambda\alpha\gamma}] - [P_\gamma, \Gamma_{\lambda\alpha\beta}]) + (\Gamma_{\lambda\beta\sigma}\Gamma^\sigma{}_{\alpha\gamma} - \Gamma_{\lambda\gamma\sigma}\Gamma^\sigma{}_{\alpha\beta}) \tag{9d}$$

where $\Gamma_{\gamma\alpha\beta}$ is to be inserted from equation (9c) for the Christoffel symbols giving an expression with only commutators. One then defines the Ricci tensor as:

$$R_{\alpha\beta} = g^{\mu\nu} R_{\alpha\mu\beta\nu} = (-i/\hbar)[P^\mu, X^\nu] R_{\alpha\mu\beta\nu} \tag{9e}$$

and also defines

$$R = g^{\alpha\beta} R_{\alpha\beta} = (-i/\hbar)[P^\alpha, X^\beta] R_{\alpha\beta} \tag{9f}$$

all of which must be inserted into the LHS of Einstein equations,

$$R_{\alpha\beta} - ½ g_{\alpha\beta} R + g_{\alpha\beta}\Lambda = (8\pi G/c^4) T_{\alpha\beta} \tag{9g}$$

Then finally we have the LHS of Einstein equations in terms of just commutators:

$$R_{\alpha\beta} + ((i/\hbar)[P_\alpha, X_\beta])(½ R - \Lambda) = (8\pi G/c^4) T_{\alpha\beta} \tag{9h}$$

where $R_{\alpha\beta}$ and $R$ are given above in terms of commutators and where $T^{\alpha\beta}$ comes from the SM as

$$T^{\alpha\beta} = \langle\Psi| \gamma^\alpha P^\beta +...|\Psi\rangle \tag{9i}$$

with the $\gamma^\alpha P^\beta$ term being symmetrized over $\alpha$ and $\beta$ and which are to act in both directions giving four terms for all fermions in the SM along with similar operator contributions from the boson fields and that of DM (discussed below) and where $|\Psi\rangle$ represents the fermions in the system and where $\langle\Psi|$ is understood to be the adjoint (complex conjugate times $\gamma^0$). With these substitutions, one obtains a very large number of terms (commutators) on the left hand side of the Einstein equation when the equations are expressed totally in terms of the commutators of $P^\mu$.

We do not need to expressly write out the SU(3) x SU(2) x U(1) gauge group as this is well developed in the literature as well as the commutators of all of these observables with the inversions ($I_s$, $I_t$, and $I_c$). When $\hbar$ is infinitesimally small compared to the parameters of the problem, then one obtains the traditional GR equations in the Heisenberg representation for the operators and particle trajectories are geodesics as follows from Einstein's equations. Likewise, when masses and their associated gravitational fields are small compared to the other parameters in the problem, then gravitation can be ignored and one obtains the standard Minkowsky metric of EP with the SM. Thus the current formulation smoothly contains QT, and GR.

## 5. Proposed Integration of the SM Gauge Groups with the EPE Algebra

The principle of minimal electromagnetic interaction states that the four-momentum $P^\mu$ is to be replaced by $D^\mu = P^\mu - e A^\mu(X)$ where $A^\mu(X)$ is the electromagnetic four-vector potential (which is a function of space-time position) and e is the electromagnetic coupling constant. When used in the Lagrangian for a Dirac spin ½ charged particle then $<\Psi| \gamma_\mu P^\mu | \Psi>$ is changed to $<\Psi|\gamma_\mu D^\mu|\Psi>$ and the Dirac equation for a particle in an electromagnetic field becomes $(\gamma_\mu(P^\mu - e A^\mu)-m)|\Psi> = 0$. This framework is invariant under the Abelian gauge transformations where $|\Psi> \rightarrow e^{i\Lambda(X)} |\Psi>$ and where $A^\mu \rightarrow A^\mu - \partial^\mu \Lambda(x)$. The requirement of invariance under these gauge transformations was the precursor to the Yang Mills introduction of the invariance of the fundamental physical observables under non-Abelian gauge transformations which evolved into the current standard model [2,3] with U(1) x SU(2) x SU(3).

We now propose (1) to further generalize the EPE algebra by replacing $P^\mu$ with the full $D^\mu$ that incorporates the mediating vector fields that serve as a representation space for the gauge groups of the standard model. Thus we now postulate that $[P^\mu, X^\nu] = i\hbar \, I \, g^{\mu\nu}(X)$ be altered to represent the effective four-momentum of the standard model as

$$[D^\mu, X^\nu] = i\hbar \, I \, g^{\mu\nu}(X) \tag{10a}$$

so we can now write

$$g^{\mu\nu}(X) = (-i/\hbar) [D^\mu, X^\nu] \tag{10b}$$

where in the position representation $|y>$, where

$$<y| D_\nu = (i\hbar (\partial/\partial y^\nu) + A_\nu) <y| \tag{10c}$$

where we now define $A^\mu$ to include all vector gauge fields as:

$$<y| A^\mu(X) = A^\mu(y) <y| = (-ig_s G_a^\mu T^a - (1/2)g' Y_w B^\mu - (1/2)g \tau_L W^\mu) <y| \tag{10d}$$

incorporating all of the vector bosons into the fundamental 15 parameter modified EPE algebra. In the quantum chromodynamics (QCD) sector the SU(3) symmetry is generated by $T^a$ while $G_a^\mu$ is the SU(3) gauge field containing the gluons. These act only on the quark fields with the strong coupling constant $g_s$. In the electroweak sector $B^\mu$ is the U(1) gauge field and $Y_w$ is the weak hypercharge generating the U(1) group while $W^\mu$ is the three component SU(2) gauge field and $\tau_L$ are the Pauli matrices which generate the SU(2) group (where the subscript L indicates that they only act on left fermions). The coupling constants are $g'$ and $g$.

We also define the SM fields as:

$$F^{\mu\nu}(X) = [D^\mu, D^\nu] = [P^\mu, P^\nu] + [P^\mu, A^\nu(X)] + [A^\mu(X), P^\nu] + [A^\mu(X), A^\nu(X)] \tag{10e}$$

as the SM generalization of the electromagnetic field tensor to encompass all intermediate vector bosons and gravitational contributions. One recalls from above that the $[P^\mu, P^\nu]$ commutator involves the derivatives of the metric tensor. This is because the covariant derivatives in Riemannian geometry do not commute when acting upon a vector field but instead give the Riemannian tensor contracted with the vector indices. $F^{\mu\nu}(X)$ now contains all of the commutators of the standard model vector bosons including both the strong and electroweak interactions. All of the gauge indices are contracted and thus do not appear explicitly in $F^{\mu\nu}(X)$. We next require, as is standard, that (2) the system be invariant under the associated strong and electroweak gauge transformations of the SM which then requires that the Dirac fields upon which this algebra is to be realized, must also undergo the corresponding gauge transformation that insures system invariance (or conversely).

We can now use [10b] and [10c] to write:
The Christoffel symbols

$$\Gamma_{\gamma\alpha\beta} = (½) (\partial_\beta \, g_{\gamma\alpha} + \partial_\alpha \, g_{\gamma\beta} - \partial_\gamma \, g_{\alpha\beta}) \tag{11a}$$

in terms of the commutators of the effective four-momentum with the metric as

$$\Gamma_{\gamma\alpha\beta} = (½) (-i/\hbar) ([D_\beta, g_{\gamma\alpha}] + [D_\alpha, g_{\gamma\beta}] - [D_\gamma, g_{\alpha\beta}]) \tag{11b}$$

Then using $g_{\alpha\beta}(X) = (-i/\hbar) [D_\alpha, X_\beta]$ one obtains

$$\Gamma_{\gamma\alpha\beta} = (-½) (1/\hbar^2) ([D_\beta, [D_\gamma, X_\alpha]] + [D_\alpha, [D_\gamma, X_\beta]] - [D_\gamma, [D_\alpha, X_\beta]]) \tag{11c}$$

The Riemann tensor becomes:

$$R_{\lambda\alpha\beta\gamma} = (-i/\hbar)\left([D_\beta, \Gamma_{\lambda\alpha\gamma}] - [D_\gamma, \Gamma_{\lambda\alpha\beta}]\right) + (\Gamma_{\lambda\beta\sigma}\Gamma^\sigma{}_{\alpha\gamma} - \Gamma_{\lambda\gamma\sigma}\Gamma^\sigma{}_{\alpha\beta}) \tag{11d}$$

where $\Gamma_{\gamma\alpha\beta}$ is to be inserted from equation (14c) for the Christoffel symbols giving an expression with only commutators. One then defines the Ricci tensor as:

$$R_{\alpha\beta} = g^{\mu\nu} R_{\alpha\mu\beta\nu} = (-i/\hbar)[D^\mu, X^\nu] R_{\alpha\mu\beta\nu} \tag{11e}$$

and also defines

$$R = g^{\alpha\beta} R_{\alpha\beta} = (-i/\hbar)[D^\alpha, X^\beta] R_{\alpha\beta} \tag{11f}$$

all of which must be inserted into the LHS of Einstein equations,

$$R_{\alpha\beta} - \tfrac{1}{2} g_{\alpha\beta} R + g_{\alpha\beta}\Lambda = (8\pi G/c^4) T_{\alpha\beta} \tag{11g}$$

Then finally we have the LHS of Einstein equations in terms of just commutators:

$$R_{\alpha\beta} + ((i/\hbar)[D_\alpha, X_\beta])(\tfrac{1}{2} R - \Lambda) = (8\pi G/c^4) T_{\alpha\beta} \tag{11h}$$

where $R_{\alpha\beta}$ and $R$ are given above in terms of commutators as shown. The form using the effective momentum D is the same as the form with P.

The following need to be recomputed using the definition $L^{\mu\nu} = X^\mu D^\nu - X^\nu D^\mu$ thus generalizing the previous equations to include both the non-commuting aspect of P and the SM generalization of P to D. Then we need to compute the spin Casimir operators and the gamma matrices. For example:

$$[L^{\mu\nu}, X^\lambda] = L^{\mu\nu} X^\lambda - X^\lambda L^{\mu\nu} \tag{12a}$$

$$= X^\mu P^\nu X^\lambda - X^\nu P^\mu X^\lambda - X^\lambda X^\mu P^\nu + X^\lambda X^\nu P^\mu$$

$$= X^\mu P^\nu X^\lambda - X^\nu P^\mu X^\lambda - X^\mu X^\lambda P^\nu + X^\nu X^\lambda P^\mu - X^\nu X^\lambda P^\mu \quad \text{since } X^\mu X^\lambda \text{ commute}$$

$$= X^\mu P^\nu X^\lambda - X^\nu P^\mu X^\lambda - X^\mu(-i\hbar\, g^{\nu\lambda}(X) + P^\nu X^\lambda) + X^\nu(-i\hbar\, g^{\mu\lambda}(X) + P^\mu X^\lambda)$$

$$= X^\mu P^\nu X^\lambda - X^\nu P^\mu X^\lambda + X^\mu i\hbar\, g^{\nu\lambda}(X) - X^\mu P^\nu X^\lambda) - X^\nu i\hbar\, g^{\mu\lambda}(X) + X^\nu P^\mu X^\lambda)$$

$$= i\hbar\,(g^{\nu\lambda}(X) X^\mu - g^{\mu\lambda}(X) X^\nu) + X^\mu P^\nu X^\lambda - X^\nu P^\mu X^\lambda - X^\mu P^\nu X^\lambda + X^\nu P^\mu X^\lambda) \text{ and as the last four terms cancel}$$

$$[L^{\mu\nu}, X^\lambda] = i\hbar\,(g^{\nu\lambda}(X) X^\mu - g^{\mu\lambda}(X) X^\nu) \text{ which is the same form as before in the standard Poincare algebra.} \tag{12b}$$

But

$$[L^{\mu\nu}, P^\lambda] = L^{\mu\nu} P^\lambda - P^\lambda L^{\mu\nu} \quad P^\lambda X^\mu = (i\hbar\, g^{\lambda\mu}(X) + X^\mu P^\lambda) \quad P^\lambda X^\nu = (i\hbar\, g^{\lambda\nu}(X) + X^\nu P^\lambda) \tag{12c}$$

$$= X^\mu P^\nu P^\lambda - X^\nu P^\mu P^\lambda - (i\hbar\, g^{\lambda\mu}(X) + X^\mu P^\lambda) P^\nu + (i\hbar\, g^{\lambda\nu}(X) + X^\nu P^\lambda) P^\mu$$

$$= X^\mu P^\nu P^\lambda - X^\nu P^\mu P^\lambda - i\hbar\, g^{\lambda\mu}(X) P^\nu - X^\mu P^\lambda P^\nu + i\hbar\, g^{\lambda\nu}(X) P^\mu + X^\nu P^\lambda P^\mu$$

$$= X^\mu P^\nu P^\lambda - X^\nu P^\mu P^\lambda - i\hbar\, g^{\lambda\mu}(X) P^\nu - X^\mu P^\lambda P^\nu + i\hbar\, g^{\lambda\nu}(X) P^\mu + X^\nu P^\lambda P^\mu$$

And with $[P^\mu, P^\nu] = i\hbar\, B^{\mu\nu}{}_\gamma P^\gamma$ thus $P^\mu P^\nu = P^\nu P^\mu + i\hbar\, B^{\mu\nu}{}_\gamma P^\gamma$ in the last two terms

$$P^\lambda P^\mu = (P^\mu P^\lambda + i\hbar\, B^{\lambda\mu}{}_\gamma P^\gamma) \quad P^\lambda P^\nu = (P^\nu P^\lambda + i\hbar\, B^{\lambda\nu}{}_\gamma P^\gamma)$$

$$= i\hbar\,(g^{\lambda\nu}(X) P^\mu - g^{\lambda\mu}(X) P^\nu) + X^\mu P^\nu P^\lambda - X^\nu P^\mu P^\lambda - X^\mu P^\lambda P^\nu + X^\nu P^\lambda P^\mu \quad \text{we get}$$

$$= i\hbar\,(g^{\lambda\nu}(X) P^\mu - g^{\lambda\mu}(X) P^\nu) + X^\mu P^\nu P^\lambda - X^\nu P^\mu P^\lambda - X^\mu(P^\nu P^\lambda + i\hbar\, B^{\lambda\nu}{}_\gamma P^\gamma) + X^\nu(P^\mu P^\lambda + i\hbar\, B^{\lambda\mu}{}_\gamma P^\gamma)$$

$$= i\hbar\,(g^{\lambda\nu}(X) P^\mu - g^{\lambda\mu}(X) P^\nu) + X^\mu P^\nu P^\lambda - X^\nu P^\mu P^\lambda - X^\mu P^\nu P^\lambda - i\hbar\, X^\mu B^{\lambda\nu}{}_\gamma P^\gamma) + X^\nu P^\mu P^\lambda + i\hbar\, X^\nu B^{\lambda\mu}{}_\gamma P^\gamma)$$

$$= i\hbar\,(g^{\lambda\nu}(X) P^\mu - g^{\lambda\mu}(X) P^\nu) - i\hbar\, X^\mu B^{\lambda\nu}{}_\gamma P^\gamma + i\hbar\, X^\nu B^{\lambda\mu}{}_\gamma P^\gamma \text{ and since } [X, B] = 0 \text{ we can simplify as}$$

$$= i\hbar\,(g^{\lambda\nu}(X) P^\mu - g^{\lambda\mu}(X) P^\nu) - i\hbar\,(B^{\lambda\nu}{}_\gamma X^\mu - B^{\lambda\mu}{}_\gamma X^\nu) P^\gamma$$

$$[L^{\mu\nu}, P^\lambda] = i\hbar\,(g^{\lambda\nu}(X) P^\mu - g^{\lambda\mu}(X) P^\nu) - i\hbar\,(B^{\lambda\nu}{}_\gamma X^\mu - B^{\lambda\mu}{}_\gamma X^\nu) P^\gamma \tag{12d}$$

which is much more complex and where the last term is new.

This has an additional pair of terms for the rotation and Lorentz transformation of $P^\lambda$ from the noncommutativity of P. This means that the Lorentz group acts differently on X and on P due to the vector boson fields in D. One can execute a translation:

$$X'^\lambda = \mathrm{Exp}((-i/\hbar)a_\mu P^\mu)\, X^\lambda\, \mathrm{Exp}((i/\hbar)a_\mu P^\mu) \tag{12e}$$

$$= (1 + (-i/\hbar)a_\mu P^\mu)\, X^\lambda\, (1 + (i/\hbar)a_\mu P^\mu)$$

$$= X^\lambda + (-i/\hbar)a_\mu [P^\mu, X^\lambda] + \text{higher order as a is infinitesimal}$$

$$= X^\lambda + (-i/\hbar)(i\hbar) g^{\mu\lambda} a_\mu$$
$$X'^\lambda = X^\lambda + a^\lambda \quad \text{which is a standard translation.} \tag{12f}$$

A New Gravitational Force appears similar to the EM fields:

From $[P^\mu, P^\nu] = i\hbar B^{\mu\nu}{}_\gamma P^\gamma$ where we define (12g)

$$B^{\mu\nu}{}_\gamma = ( g^{\mu\alpha}(y) (\partial g^{\upsilon\beta}(y)/\partial y^\alpha) - g^{\nu\alpha}(y) (\partial g^{\mu\beta}(y)/\partial y^\alpha) ) g_{\beta\gamma}(y) \tag{12h}$$

Whose origin is in the asymmetry of the derivatives of the metric

$B^{\mu\nu} = B^{\mu\nu}{}_\gamma P^\gamma$ is the gravitational analogue of $F^{\mu\nu}$ where (12i)

$$B^{\mu\nu}{}_\gamma = g^{\mu\alpha} g_{\beta\gamma} \partial g^{\upsilon\beta}/\partial y^\alpha - g^{\nu\alpha} g_{\beta\gamma} \partial g^{\mu\beta}/\partial y^\alpha \tag{12j}$$

Understanding that g is a function of the position one can simplify to

$$B^{\mu\nu} = ( g^{\mu\alpha} \partial g^{\upsilon\beta}/\partial y^\alpha - g^{\nu\alpha} \partial g^{\mu\beta}/\partial y^\alpha ) P_\beta \tag{12k}$$

which is like a new gravitational force:

$$B^{\mu\nu} = ( \partial^\mu g^{\upsilon\beta} - \partial^\nu g^{\mu\beta} ) P_\beta \tag{12l}$$

This "force" arises naturally as an addition to the forces due to the gauge fields of the vector bosons. As it originates from the Riemann metric, it might seem be be related to dark energy as it is dependent upon the metric and its derivatives. But there is no adjustable parameter. Again this "force" term comes from the fact that covariant derivatives do not commute when acting upon a vector field but can be written in terms of a contraction of the field with the Riemann tensor.

We will call the EPE algebra that has been extended by $D^\mu$ to encompass the SM as the EPESM algebra. All other commutators in the EPESM algebra must now be studied including those for angular momentum, spin and spin tensors (including the $\gamma$ matrices). These computations are underway but are complicated especially with the inclusion of the three inversion operations which span both the EPE and SM algebras. The resulting EPESM algebra contains both the 15 parameter EPE algebra of the space-time variables and the 12 parameter Lie algebra of the standard model. This algebra must also support the 3 parameter algebra of inversions ($I_t$, $I_s$, $I_c$) giving an algebra with a total of 30 observables.

The customary SM Lagrangian can now be written in terms of the algebra's representations (fields) with the customary dynamics based upon Feynman paths with the metric tensor determined from Einstein's equations now framed in algebraic form. But the energy-momentum tensor is more complex being of the form $T^{\mu\nu} = <\Psi|\gamma^\mu D^\nu|\Psi>$ symmetrized and acting in both directions with $D^\nu$ containing all vector bosons. It is understood that $<\Psi|$ represents the Dirac adjoint (complex conjugate times $\gamma^0$) with separate components for the quarks and leptons. There are additional terms for the Higgs complex scalar field under the SU(2) group for the Lagrangian that have not been covered here.

## 6. Discussion and Conclusions:

Our first integration postulate (1) was that the Heisenberg structure constants, with the Minkowsky metric $g^{\mu\nu}$, be generalized to be functions of the four-vector position operators whose values were to be determined by Einstein's GR equations for the metric (in the position representation) from the energy momentum tensor with Einstein's differential equations for the Christoffel symbols and the Riemann and Ricci tensors restated as commutation rules. This "Extended Poincare Einstein" (EPE) algebra is like a Lie algebra but with structure constants that vary over space and time as determined by Einstein's equations for general relativity (GR). As such, it represents a new mathematical structure where unlike a gauge transformation based algebra one here has a different Lie algebra at different points in space time.

Our second postulate was to integrate the SM with the EPE algebra by generalizing $P^\mu$ to be the effective four momentum, $D^\mu$, which includes the gauge vector bosons with the requirement of invariance under the SU(3) x SU(2) x U(1) gauge transformations of the SM resulting in an algebraic commutator structure involving 15 space-time observables, 12 gauge observables, and the three inversions. We called this 30 parameter algebra the EPESM algebra to indicate the integration of the SM gauge groups. The admissible particles (fields) are to be representations of this 30 parameter algebra of observables and particle interactions are to result from the SM Lagrangian with a Feynman path integral dynamics. The resulting algebraic structure involves structure "constants" that vary over space-time due to the gravitationally altered metric tensor and also involves the SM non-Abelian gauge transformations that involve both the spin ½ quarks and leptons canceling the gauge transformations of the vector bosons that mediate the strong and electroweak force. These two postulates lay a foundational structure for our proposed means of integration.

Yet this work leaves unanswered questions: 1. The full commutation rules for the Lorentz component of the extended Poincare algebra including those for angular momentum, spin, the associated Casimir operators and the gamma matrices need to be derived as a result of both cases from postulates I (GR) and II (SM) using the standard definitions. 2. Then to construct and solve the associated Dirac equation for a hydrogen atom, and harmonic oscillator for simple molecules in order to test whether this approach has observable testable consequences in strong gravitational fields such atomic or molecular spectra as near black holes. Specifically one seeks some prediction that would only be consequential to this new formulation and not one that could be obtained by simply using known aspects of classical GR. 3. Another task is to develop the associated equations and interpretation for parallel transport, geodesics, and a general framework and interpretation for general coordinate transformations to formulate the full invariance of the theory but now using X, P, and D operators rather than the traditional Riemannian geometry. This must especially address the complexities of the differences in these operators. 4. Of particular interest is the noncommutativity of the four momentum which leads to what seems to be an additional gravitational effect as described in equations (15g) – (15l) which can be reexpressed in terms of the Riemann tensor.

It is not clear how to find all representations or even local representations of this algebra. For cases where the GR equations can be solved for the metric, then this metric gives the structure constants of the EPE algebra and thus from these one can create the regular representations. Also for such EPE solutions one can form the Cartan-Killing inner product to study the structure of the EPE. The solutions here to this algebraic structure do not yet seem easier than solving Einstein's equations for the metric due to various configurations.

We note four aspects of this proposal: (1) Our design accepts the currently standard versions of the Poincare and Heisenberg Lie Algebras (except for generalizing the structure constants), the SM, Einstein's GR, and the algebra of the three discrete inversions. (2) c, ℏ, and G are all in the structure constants of this EPE algebra on an equal footing and as such define the natural scale for mass, length and time commonly known as the Plank scale. (3) Specifically all nonzero EPE structure constants contain the metric (and its derivatives) for a curved space-time, $i\hbar\, g^{\mu\upsilon}(X)$. And (4) general relativity as formulated here is not an explicit part of the phenomenological SM of interactions but rather is a rich extension (EPE) of the kinematic space-time infrastructure of QT whose representations are to support the allowable particles in the SM which give the complementary formulation of the strong and electroweak forces. (5) Our approach is in keeping with the principles of Mach and Einstein that gravitation is a result of the Riemann curvature of space-time and not a separate force to be incorporated like other forces in the SM. More specifically, we are suggesting that gravitation manifests itself through the structure constants of the fundamental observables. (5) The uncertainty relation: $\Delta x^\mu\, \Delta p^\mu \geq \hbar\, g^{\mu\upsilon}/2$ is modified so that the metric now alters the effective value of ℏ both for the position-momentum and the energy-time inequalities. (6) The energy-momentum tensor operators in Einstein's equations are for the particles/fields (energy and momentum) found in nature and thus this $T_{lj}$ operator must originate in the SM formalism. The commutator (algebraic) form of $T^{\mu\upsilon}$ is not totally clear in terms of the Belafonte tensor and other known problems with the energy momentum tensor in its most general form. (7) One also notes that this design we are proposing in terms of just commutator rules, structure constants and algebras is in exact accord with Sophius Lie's original concept of inventing Lie algebras for the study of differential equations.

Currently the λ term in Einstein's equation is the simplest explanation for dark energy (DE) and is here a pure manifestation of the GR formulation of the EPE structure constants. Of course the question still remains of how this term arises. It currently appears that dark matter (DM) may be a new type of particle that has no strong, electromagnetic, or weak interactions but only gravitational. If this is true, then it would not necessarily emerge in any natural way from the current SM which was built on these interactions but could be a particle that carried mass yet had only gravitational interactions. Thus it would be just another representation of the EPE algebra that needed to be adjoined to the SM framework. Current experiments do not require that DM is a particle that is currently represented in the SM. Thus DM can be just another particle that only has a gravitational interaction just like the leptons do not have strong interactions.. Following the current successful methodology of the gauge transformations with the SM, there could even be another gauge group just for particles that only interact gravitationally thus extending the 12 parameter SM gauge group. If there is a stable DM particle that is part of the SM but which only has gravitational interactions, it may be difficult to measure its mass and possible spin. Now that one has the state of the system fixed, the system dynamics would be determined by using the method of Feynman path integrals using the SM Lagrangian. Although the SM Lie algebra of internal quantum observables and interactions is known, it is still a work in progress with a large number of arbitrary parameters. Also since the SM contains observables from the EPE algebra, it is necessary to cast the SM framework into an algebraic structure where derivatives are replaced by commutators and gamma matrices by spin representations. One notes that our approach fully integrates GR with spin and angular momentum as well as space-time. Then one can ask if the resulting algebraic system self-consistent when the GR equations are invoked between T and g.

By using those cases where Einstein's equations have been solved, one can determine $g^{\mu\nu}(X)$ and thus can explicitly write the structure constants for the EPE and seek representations of that algebra in neighborhoods where the metric can be considered locally constant. Thus one begins with the state of a physical system (such as a black hole with some particle state exterior to it such as a nucleus, atom, or molecule) some distance outside the event horizon. Then the energy momentum tensor can be classically determined at the location of the object and Einstein's equations for the metric tensor can be solved for this case. Then that metric tensor can be inserted into the structure constants for the EPE algebra. One must then find the representations of that local Lie algebra which give the allowable states in nature which are then required to support the SM gauge groups along with the resulting altered Dirac equation. The immediate objective is to seek new predictions that issue from this EPE algebraic framework.

In particular for a spherical mass, the Schwarzschild metric has

$g_{00} = (1- r_s /r)$ and (13a)

$g_{rr} = -1/(1- r_s /r)$ where the Schwarzschild radius is given by (13b)

$r_s = 2GM/c^2$ (13c)

and r is the radius of the mass M located at r = 0 as given in spherical coordinates. This implies that $\Delta X\, \Delta P$ and $\Delta t\, \Delta E$ have effectively different values from traditional quantum theory if the gravitational field is very large such as near a black hole. For the case of a non-rotating black hole without charge we get

$\Delta X_r \Delta P_r \geq (\hbar/2)(1/(1-r_s/r))$ and (13d)

$\Delta t \Delta E \geq (\hbar/2)(1-r_s/r)$ (13e)

which modify the uncertainty principle in a strong gravitational field. It follows that taking the product of these two equations that one gets:

$\Delta X_r \Delta P_r\, \Delta t \Delta E \geq (\hbar/2)^2$ which is independent of both r and $r_s$ (13f)

It is not obvious how to measure these altered uncertainty relations so one needs to seek effects that would be sensitive to these altered states of "effective ℏ" due to the metric, as might be observable in either energy transitions or angular momentum values. For example, the Dirac equation for hydrogen in a strong gravitational field has the time and $X^3 = r$ component modified by these Schwarzschild metric factors so the energy levels would be altered with modified emission spectra, possibly observable. Here one can take the radial direction from the center of the black hole to be the $X^3$ direction since the black hole field appears as perpendicular to a plane from a position near the event horizon while the x and y directions are not affected as the black hole is so effectively large at this proximity. The Dirac equation for the states of the hydrogen atom thus need to be solved thus giving an asymmetric Dirac equation to be solved for the energy levels and transitions. In a subsequent article we will address the exact commutation relations for the remaining observables in the EPESM algebra and other possible observable consequences of these assumptions.

A particle would here move along Feynman paths near a geodesic and that this geodesic would be altered by the strong and electroweak forces implied by the SM. This suggests the development of a Foldy-Wouthuysen type formulation that generalizes the motion of a charged Dirac particle in an electromagnetic field [10,2] where the free particle would evolve close to the geodesic along the Feynman paths that would create the least phase interference. The SM formulation would then cause deviations from that geodesic as required by the strong and electroweak interactions. The dynamical evolution of the entire system would follow current methodology using Feynman paths as this provides the required covariant treatment of time.

The framework that we are seeking here is an algebra of fundamental operators $L_i$ that represent actions or measurements that can be taken on the physical world and defined by their commutation rules which exactly defines their mutual interference of action. Then the entities (particles, fields, quanta) that make up the physical world are exactly the representations of that algebra on a representation space with a scalar product ( $a$, $a^+$ acting on the vacuum with known commutation rules) that give the probability amplitudes for processes. The question is whether the framework presented here can evolve with the changing standard model, be shown to be self-consistent, and offer predictions for new phenomena.